\documentclass{ws-procs975x65}

\def\beq{\begin{equation}}
\def\eeq{\end{equation}}

\begin{document}

\title{Progenitors of ultra-stripped supernovae}
\author{Thomas M. Tauris$^*$, Norbert Langer}
\address{AIfA, University of Bonn / MPIfR\\
53121 Bonn, Germany\\
$^*$E-mail: tauris@astro.uni-bonn.de}
\author{Philipp Podsiadlowski}
\address{Department of Astronomy, Oxford University,\\
Oxford OX1 3RH, UK}

\begin{abstract}
The explosion of ultra-stripped stars in close binaries may explain new discoveries of weak and fast optical transients. 
We have demonstrated that helium star companions to neutron stars (NSs) may evolve into naked 
metal cores as low as $\sim\!1.5\;M_{\odot}$, barely above the Chandrasekhar mass limit, by the time they explode. Here we present a new systematic 
investigation of the progenitor evolution leading to such ultra-stripped supernovae (SNe), in some cases yielding pre-SN envelopes 
of less than $0.01\;M_{\odot}$. We discuss the nature of these SNe (electron-capture vs iron core-collapse) and their observational 
light-curve properties. Ultra-stripped SNe are highly relevant for binary pulsars, as well as gravitational wave detection of merging NSs by LIGO/VIRGO, 
since these events are expected to produce mainly low-kick NSs in the mass range $1.10-1.80\;M_{\odot}$.
\end{abstract}


\bodymatter


\section{Final Stage of Mass Transfer prior to the SN Explosion}
\begin{figure}[t]
\begin{center}
\includegraphics[width=12.5cm, angle=0]{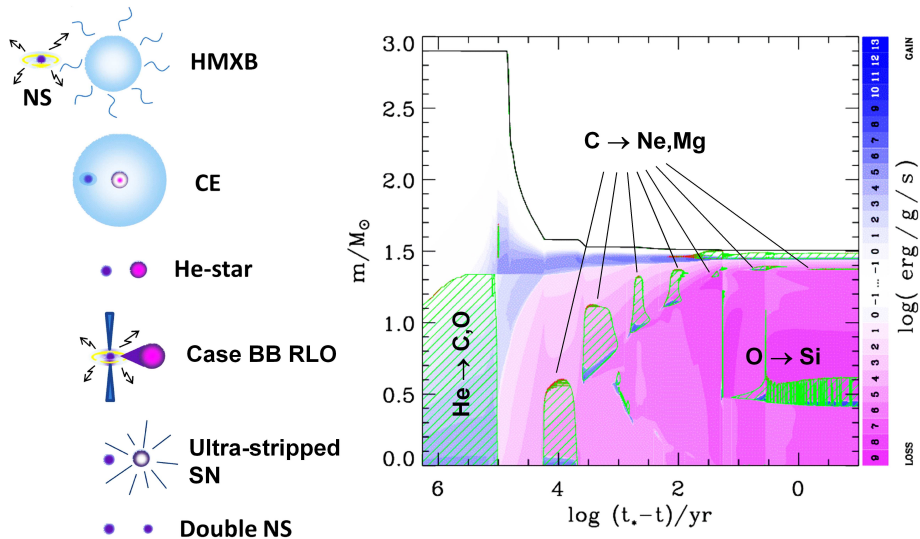}
\end{center}
\caption{Left: cartoon of evolutionary sequence of a binary from a HMXB to a DNS system. Right: Kippenhahn diagram
(stellar cross-section in mass coordinates versus remaining calculated lifetime) of a $2.9\;M_{\odot}$ helium star losing mass via Case~BB RLO. 
At the end of the mass transfer it has become an almost naked metal core with a total mass of $\sim\!1.50\;M_{\odot}$. 
The plot shows the evolution until about 10~years prior to an ultra-stripped Fe~CCSN \cite{tlm+13}.
 }
\label{fig:Kippenhahn}
\end{figure}

To form a double neutron star (DNS) system, a pair of massive stars have to evolve through
(and survive) a long chain of binary interactions \cite{tv06,lan12}, including mass transfer and supernova (SN) explosions.
Following the first SN, a bound system evolves to become a high-mass X-ray binary (HMXB) in which a NS accretes material from the wind of its massive companion star
(or accretes material accumulated from the passage through the decretion disk of a rapidly rotating B-star, i.e. a Be/X-ray binary).
Once the companion star fills its Roche lobe, the binary system becomes dynamically unstable on a very short timescale of a few hundred years \cite{sav78},
leading to the formation of a common envelope (CE) \cite{pac76}. Depending on the binding energy of the envelope and the amount of liberated
orbital energy, the system may avoid a merger and survive as an exposed core, a naked helium star, orbiting the NS.
Since helium stars expand during their giant phase (when undergoing helium shell burning), they may give rise to
an additional epoch of Roche-lobe overflow (RLO), i.e. so-called Case~BB RLO \cite{st76,dd77,dt81,hab86a}, cf. Fig.~\ref{fig:Kippenhahn}.
During this stage, the donor star becomes significantly stripped prior to its core collapse. As a result of the small amount of
ejecta mass during the subsequent SN, possibly accompanied by a small NS kick, the binary is likely to remain bound
and leave behind a DNS system.

There are currently 15 DNS systems known, including a few unconfirmed candidates. We have recently argued \cite{tlp15}
that possibly all DNS systems evolved via Case~BB RLO prior to the core collapse producing the second NS in such a system.

\section{Ultra-stripped SNe}
\begin{figure}[t]
\begin{center}
\includegraphics[trim=0 0 0 -1cm, width=7.0cm, angle=-90]{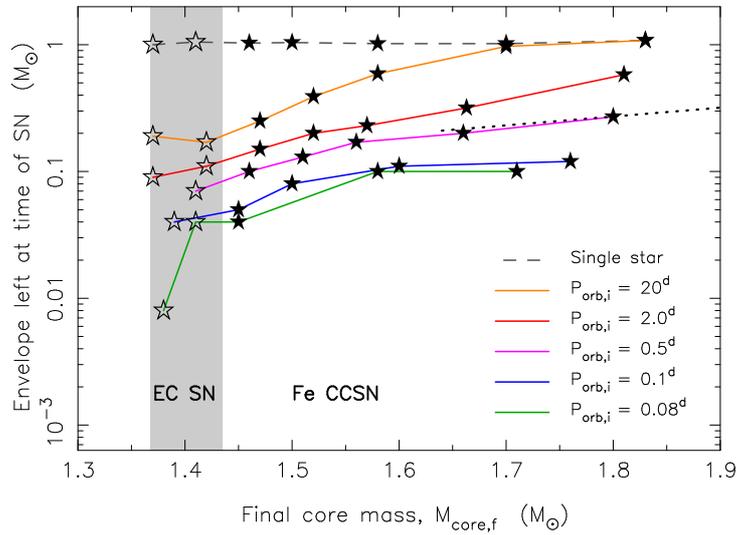}
\end{center}
\caption{Total remaining envelope mass versus final core mass prior to SN explosion. At the onset of Case~BB RLO, the NS companion was a helium star with
a mass between $2.6-3.5\;M_{\odot}$ and an orbital period between $0.08-20\;{\rm days}$. Open stars indicate a final fate as an EC~SN and solid stars
indicate an Fe~CCSN. The dashed line at the top shows calculations for single helium stars. The dotted line shows the minimum envelope mass of cores
following RLO to a main-sequence star \cite{ywl10}. This figure was adopted from Ref.~[5].
}
\label{fig:env_mass}
\end{figure}

\begin{figure}[h]
\begin{center}
\includegraphics[trim=0 0 0 0, width=7.0cm, angle=-90]{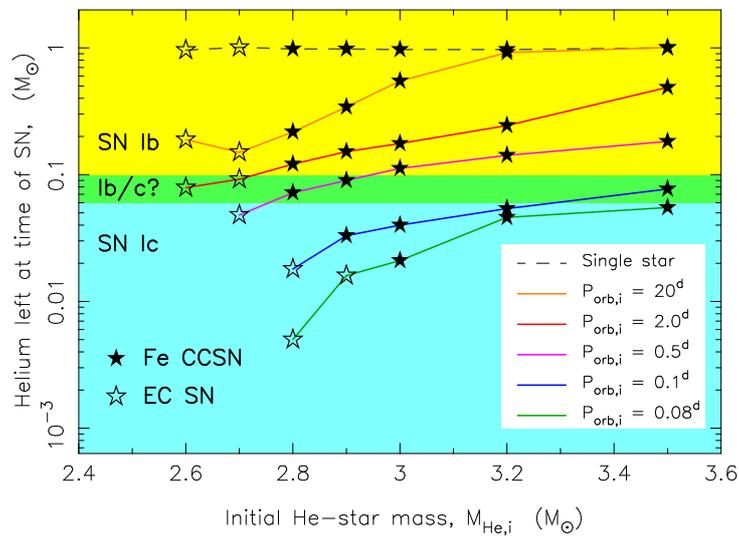}
\end{center}
\caption{Total amount of helium prior to the explosion, and the expected classification of observed SNe as Type~Ib or Ic, for the same models shown in Fig.~\ref{fig:env_mass}. 
This figure was adopted from Ref.~[5].}
\label{fig:TypeSN}
\end{figure}

As a consequence of the compact nature of the NS, it is able to strip (almost) the entire helium envelope of its companion star while
this star evolves through helium, carbon, oxygen and silicon burning, and loses mass to the NS via Case~BB RLO at the same time.
Depending on the initial orbital period and mass of the helium star at the onset of Case~BB RLO, the final metal core
left behind prior to core collapse often has an envelope mass of $\le0.1\;M_{\odot}$ -- in extreme cases even $<0.01\;M_{\odot}$ \cite{tlp15}, 
cf. Figs.~\ref{fig:env_mass} and \ref{fig:TypeSN}.
The explosion of such a star thus results in an {\it ultra-stripped SN}\, \cite{tlm+13,tlp15}.

We find that the nature of the resulting ultra-stripped core collapse can be either an electron-capture SN \cite{plp+04} (EC~SN) or an iron core-collapse SN (Fe~CCSN) 
if the metal core mass is between $\sim1.37-1.43\;M_{\odot}$ or above $\sim1.44\;M_{\odot}$, respectively.
\begin{figure}[t]
\begin{center}
\includegraphics[trim=0 0 0 -1cm, width=8.0cm, angle=-90]{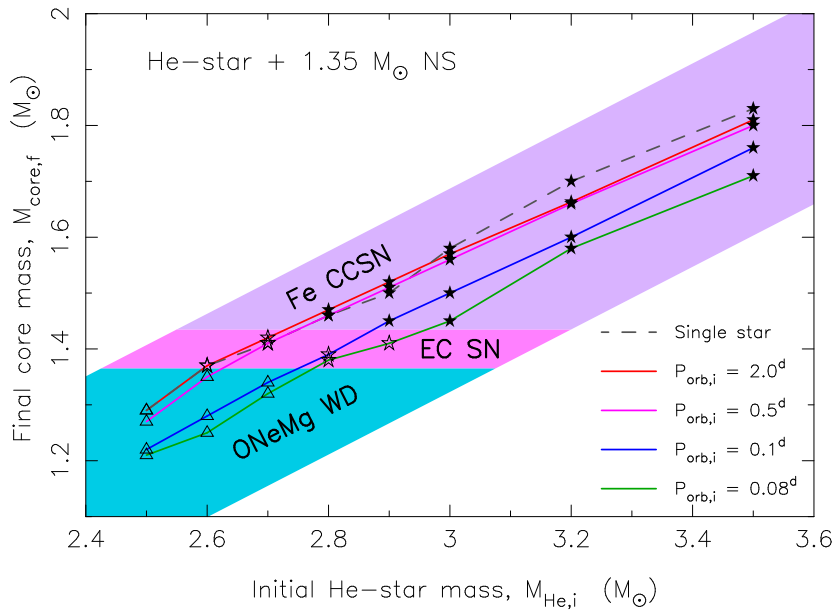}
\end{center}
\caption{Sequences of final core mass as a function of initial helium star mass, for different values of initial orbital period prior to Case~BB RLO.
The final destiny of these helium stars is either an Fe~CCSN (purple region, filled stars), an EC~SN (pink region, open stars), or an ONeMg~WD (blue region, triangles).
The symbols connected by a dashed line were calculated for an isolated helium star. In all other cases shown, the helium star was evolved in a binary 
with a $1.35\;M_{\odot}$ NS. This figure was adopted from Ref.~[5].
}
\label{fig:Mcore_MHe}
\end{figure}
In Fig.~\ref{fig:Mcore_MHe} we show the final destiny for our calculated systems consisting of a $1.35\;M_{\odot}$ NS orbiting a helium star with an initial mass
of  $M_{\rm He}=2.5-3.5\;M_{\odot}$ and orbital period between 0.08 and $2.0\;{\rm days}$. EC~SNe are only possible within a relatively 
narrow mass interval of $M_{\rm He}=2.60-2.95\;M_{\odot}$, depending the initial orbital period. Below and above this mass interval the outcome is
an ONeMg white dwarf (WD) or an Fe~CCSN, respectively. All abovementioned explosions will be ultra-stripped SNe, as a consequence of severe stripping 
of the exploding star via Case~BB RLO.

\section{Observational Signatures of Ultra-stripped SNe}
Depending on the amount of helium in the ejected envelope, and the amount of nickel
mixed into this ejecta, we expect that these ultra-stripped SNe can be classified as both Type~Ib or Type~Ic, cf. Fig.~\ref{fig:TypeSN}.
Recent observations of SN~2005ek \cite{dsm+13} indicate that this Type~Ic SN represents the smallest ratio ever detected between ejecta mass and remnant mass.
According to SN light curve modelling \cite{tlm+13}, SN~2005ek could indeed be the first observed case of an ultra-stripped SN.
Hopefully, near future high-cadence surveys and dedicated SN searches with will discover more modest luminosity SNe
with similar rapidly decaying light curves ($\Delta m_{15}\ge 3.5$). 

\begin{figure}[t]
\begin{center}
\includegraphics[trim=0 0 0 -1cm, width=8.0cm, angle=-90]{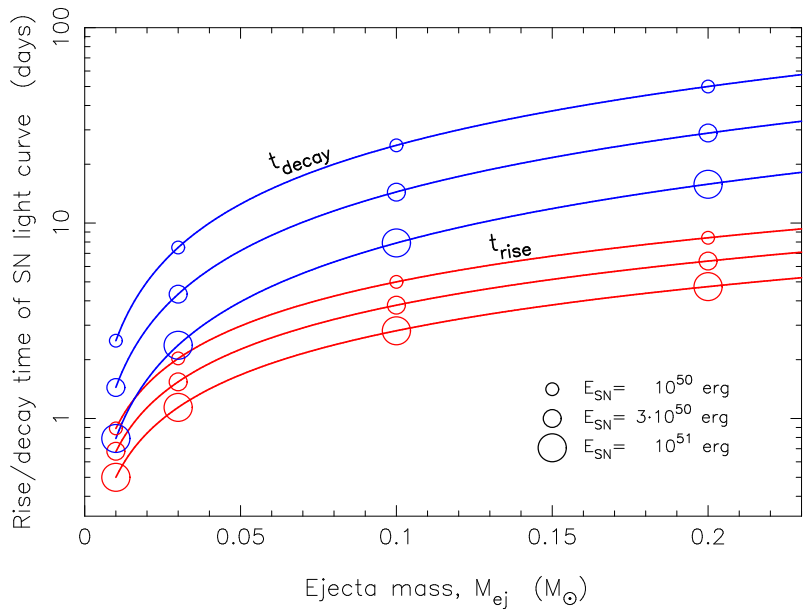}
\end{center}
\caption{Rise times (lower three curves, red) and decay times (upper three curves, blue) of light curves expected from ultra-stripped SNe with different explosion energies
               between $10^{50}-10^{51}\;{\rm erg}$.}
\label{fig:LC}
\end{figure}

The expected light-curve properties of ultra-stripped SNe can be estimated from the photon diffusion time through a homologously expanding SN envelope \cite{arn79,arn82,kk14}
and we estimate the rise time and the decay time as \cite{tlp15}:
\begin{equation}
  \tau _{\rm rise}=5.0\,{\rm d}\;\,M_{0.1}^{3/4}\;\kappa_{0.1}^{1/2}\;E_{50}^{-1/4} \qquad\qquad
  \tau _{\rm decay}=25\,{\rm d}\;\,M_{0.1}\;\kappa_{0.1}^{1/2}\;E_{50}^{-1/2}
\end{equation}  
where $M_{0.1}$ is the ejecta mass in units of $0.1\;M_{\odot}$, $E_{50}$ is the SN kinetic energy in units of $10^{50}\;{\rm erg}$, and $\kappa_{0.1}$ is the opacity
in units of $0.1\;{\rm cm}^2\,{\rm g}^{-1}$. Graphical representations of these light-curve timescales are plotted in Fig.~\ref{fig:LC}.

\section{Rate of Ultra-stripped SNe and Merging DNS Systems}
We expect \cite{tlp15} ultra-stripped SNe to leave behind NSs with masses between $1.10-1.80\;M_{\odot}$, often accompanied by small kicks, in DNS systems. 
Evidence for ultra-stripped SNe producing small kicks in general has recently been supported in an independent study of exploding bare CO stars \cite{suw15}. 
Ultra-stripped SNe are therefore highly relevant  for LIGO/VIRGO since most (probably all) merging DNS systems have evolved through such explosions.
Given that the estimated merger rate of DNS systems in a Milky Way-equivalent galaxy is of the order a few $10^{-6}-10^{-5}\;{\rm yr}^{-1}$, and the
Galactic rate of CCSNe is of the order $10^{-2}\;{\rm yr}^{-1}$, we notice that the fraction of ultra-stripped SNe to all SNe is at least 0.1~\%.
However, it is likely that a BH, and possibly even a WD, is also able to strip a Roche-lobe filling helium star prior to its SN, as effectively as an accreting NS.
In that case, explosions producing BH-NS binaries and systems containing a young pulsar orbiting a WD, will be ultra-stripped SNe as well, and thus
the total fraction of ultra-stripped SNe to all SNe could be larger -- perhaps reaching the 1~\% level.

\section{Location of Ultra-stripped SNe in Host Galaxies}
Knowing the expected location of ultra-stripped SNe with respect to star-forming regions in host galaxies will help identifying optical transients as ultra-stripped SNe.
It is known that, in average, HMXBs have small systemic velocities of the order of $10\;{\rm km\,s}^{-1}$. Such small velocities are supported by the
typical small migration distances of HMXBs from the spiral arms in the Milky Way \cite{cc13}. Hence, the travel distance of binaries giving rise to ultra-stripped SNe
is simply limited by the product of this velocity and the lifetime of the progenitor of the exploding star (typically $7-40\;{\rm Myr}$).
The resulting offsets from star forming regions in host galaxies is therefore of the order $100-400\;{\rm pc}$ (in extreme and rare cases, a maximum
distance of a few kpc might be possible -- a population synthesis study should be able to provide a distribution of offset distances).

\section{Future Work on Ultra-stripped SNe}
So far, progenitor models of helium stars leading to ultra-stripped SNe have not been evolved until the onset of core collapse.
Such computations are needed to provide input models for detailed SN explosions which may be able to determine the resulting
NS masses and kick velocities. For example, it still needs to be verified that the expected kicks associated with ultra-stripped SNe are in agreement
with constraints from observations of DNS systems. 

An other important issue to study is whether WDs are able to strip the evolved helium stars
sufficiently to produce ultra-stripped SNe too. Two close binaries are currently known (PSR~J1141$-$6545 and PSR~B2303+46) in which an old massive WD
is being orbited by young radio pulsar -- evidently the NS formed after the formation of the WD in both of these systems \cite{ts00}. 
Hence, it is likely the WD was able to significantly strip the envelope of the expanding helium star progenitor, given the relatively short
orbital periods of these two systems ($0.20\;{\rm days}$ and  $12\;{\rm days}$, respectively).

Finally, detailed modelling of the recycling and spin-up of the accreting NS during Case~BB RLO has to be investigated further.
Initial studies \cite{tlk12,tlp15} indicate that the computed spin periods of the recycled NSs in DNS  systems are indeed 
in good agreement with the observed range of $23-185\;{\rm ms}$. Work is in progress to answer this question and other
remaining issues related to ultra-stripped SNe.

\end{document}